\documentclass[%
preprint,
 amsmath,amssymb,
 aps,
prx,
]{revtex4-2}

\usepackage{graphicx}
\usepackage{bm}
\usepackage{hyperref}
\usepackage{upgreek}

\begin{document}


\title{Spectroscopic Visualization of a Robust Electronic Response of Semiconducting Nanowires to Deposition of Superconducting Islands}



\author{Jonathan Reiner}
\thanks{These two authors contributed equally}
\affiliation{Department of Condensed Matter Physics, Weizmann Institute of Science, Rehovot 7610001, Israel.}

\author{Abhay Kumar Nayak}
\thanks{These two authors contributed equally}
\affiliation{Department of Condensed Matter Physics, Weizmann Institute of Science, Rehovot 7610001, Israel.}

\author{Amit Tulchinsky}
\affiliation{Department of Condensed Matter Physics, Weizmann Institute of Science, Rehovot 7610001, Israel.}

\author{Aviram Steinbok}
\affiliation{Department of Condensed Matter Physics, Weizmann Institute of Science, Rehovot 7610001, Israel.}

\author{Tom Koren}
\affiliation{Department of Condensed Matter Physics, Weizmann Institute of Science, Rehovot 7610001, Israel.}

\author{Noam Morali}
\affiliation{Department of Condensed Matter Physics, Weizmann Institute of Science, Rehovot 7610001, Israel.}

\author{Rajib Batabyal}
\affiliation{Department of Condensed Matter Physics, Weizmann Institute of Science, Rehovot 7610001, Israel.}

\author{Jung-Hyun Kang}
\affiliation{Department of Condensed Matter Physics, Weizmann Institute of Science, Rehovot 7610001, Israel.}

\author{Nurit Avraham}
\affiliation{Department of Condensed Matter Physics, Weizmann Institute of Science, Rehovot 7610001, Israel.}

\author{Yuval Oreg}
\affiliation{Department of Condensed Matter Physics, Weizmann Institute of Science, Rehovot 7610001, Israel.}

\author{Hadas Shtrikman}
\affiliation{Department of Condensed Matter Physics, Weizmann Institute of Science, Rehovot 7610001, Israel.}

\author{Haim Beidenkopf}
\email[]{haim.beidenkopf@weizmann.ac.il}
\affiliation{Department of Condensed Matter Physics, Weizmann Institute of Science, Rehovot 7610001, Israel.}


\date{\today}

\begin{abstract}
Following significant progress in the visualization and characterization of Majorana end modes in hybrid systems of semiconducting nanowires and superconducting islands, much attention is devoted to the investigation of the electronic structure at the buried interface between the semiconductor and the superconductor. The properties of that interface and the structure of the electronic wavefunctions that occupy it determine the functionality and the topological nature of the superconducting state induced therein. Here we study this buried interface by performing spectroscopic mappings of superconducting aluminum islands epitaxially grown \textit{in-situ} on indium arsenide nanowires. We find unexpected robustness of the hybrid system as the direct contact with the aluminum islands does not lead to any change in the chemical potential of the nanowires, nor does it induce a significant band bending in their vicinity. We attribute this to the presence of surface states bound to the facets of the nanowire. Such surface states, that are present also in bare nanowires prior to aluminum deposition, pin the Fermi-level thus rendering the nanowires resilient to surface perturbations. The aluminum islands further display Coulomb blockade gaps and peaks that signify the formation of a resistive tunneling barrier at the InAs-Al interface. The extracted interface resistivity, $\rho \approx 1.3 \times 10^{-6} \ \Omega \ \mathrm{cm^2}$, will allow to proximity-induce superconductivity with negligible Coulomb blockade effects by islands with interface area as small as 0.01 $\upmu \mathrm{m^2}$. At low energies we identify a potential energy barrier that further suppresses the transmittance through the interface. A corresponding barrier exists in bare semiconductors between surface states and the accumulation layer, induced to maintain charge neutrality. Our observations elucidate the delicate interplay between the resistive nature of the InAs-Al interface and the ability to proximitize superconductivity and tune the chemical potential in semiconductor-superconductor hybrid nanowires.
\end{abstract}


\maketitle

\section{Introduction}

Metal-semiconductor interfaces, which are of prime importance for device engineering and electronic applications, have gained vast scientific attention over the past few decades \cite{monch1995semiconductor, luth2010solid}. Comprehending and controlling the properties of such interfaces is needed, for instance, for contacting or gating the semiconducting devices. More recently, the desire to induce one-dimensional (1D) topological superconductivity in semiconducting nanowires with strong spin-orbit coupling necessitated their interfacing with superconducting metals such as aluminum (Al) \cite{Lutchyn2010a, Oreg2010a, Das2012a, Mourik2012a, Lutchyn2018}. The nature and properties of the metal-semiconductor interface bears crucial consequences on the functionality of such topologically superconducting devices. Modelings of such interfaces in hybrid InAs-Al systems have mostly concluded that the introduction of metallic islands will strongly perturb the electronic occupation of the semiconducting nanowires \cite{Woods2018EffectiveDevices, Antipov2018EffectsNanowires, Mikkelsen2018HybridizationInterfaces, Escribano2019, dassarma2017dot}. The hybridization of the metallic and semiconducting states will further modify their electronic properties, as spin-orbit and Zeeman couplings, which will fundamentally affect the device character.

Bulk semiconductors are indeed known to be highly susceptible to perturbations due to their low charge density. Although the metal contact may provide improved electrostatic screening, that the semiconductor lacks, it also perturbs the electronic structure of the semiconductor due to the large work function difference between the two on the order of a few electron-volts. The resulting potential difference at the interface gives rise to a strong band bending of the semiconductor spectrum as it equilibrates with the metal \cite{Schottky1939, Mott1939}. The surface band distortion is poorly screened within the semiconductor, whose electronic occupation gradually recovers over a typical scale of several tens of nanometers, while modifying the bulk Fermi energy level as the charge redistributes. The induced charge layer gives rise to rectification in the charge transmittance across the interface, which is undesired for contacting nor for inducing superconductivity. Semiconducting nanowires were mainly modeled theoretically within this paradigm. 

However, in many cases the bare surface of the semiconductor alone acts as a strong perturbation to the bulk crystallographic order for which the semiconductor will also develop extreme electronic response. Common response to many semiconducting materials, is the formation of metallic surface states which are bound to the surface atomic layer \cite{Olsson1996a}. These can be induced as a result of various surface related perturbations including a polar surface, low impurity concentration or crystallographic irregularities, oxidation and passivisation \cite{monch1995semiconductor, luth2010solid}. The charge nature of the surface states, as acceptor or donor like, is dictated by their energy distribution relative to the bulk conduction and valence bands of the semiconducting band structure. To maintain charge neutrality, the surface states, that are bound to the atomic surface layer, are accompanied by a spatially extended charge layer of opposite charge polarity whose penetration away from the surface is set by the poor screening of the semiconductor. Accordingly, a surface dipole forms between the surface and charge layers. In turn, the semiconductor becomes fairly resilient to surface perturbations in presence of such surface states \cite{Bardeen1947SurfaceStates}. In particular, its bulk Fermi level becomes pinned at about the energy at which charge neutrality is achieved. Here we show that within our energy resolution the Fermi level of the InAs nanowires is insusceptible to deposition of metallic Al islands on their facets. Using scanning tunneling microscopy (STM) and spectroscopic mappings on the Al islands, on exposed InAs segments in their vicinity and on bare nanowires, we find evidence of Fermi level pinning and of the existence of a tunneling barrier at the interface which we attribute to the surface dipole. Our findings allude to the presence of dense surface states on the side facets of InAs nanowires which are weakly affected by the incorporation of Al. This calls for a renewed analysis of proximity induced superconductivity and the effective system parameters, such as the Land\'{e} factor and spin-orbit coupling, in hybrid InAs/Al nanowires. Our insights offer previously overlooked potential advantages, as improved robustness of the InAs/Al system, alongside new challenges that need to be addressed to further promote it as a platform for topological superconductivity applications.

\section{Results}

\begin{figure}
\includegraphics[width=1\linewidth]{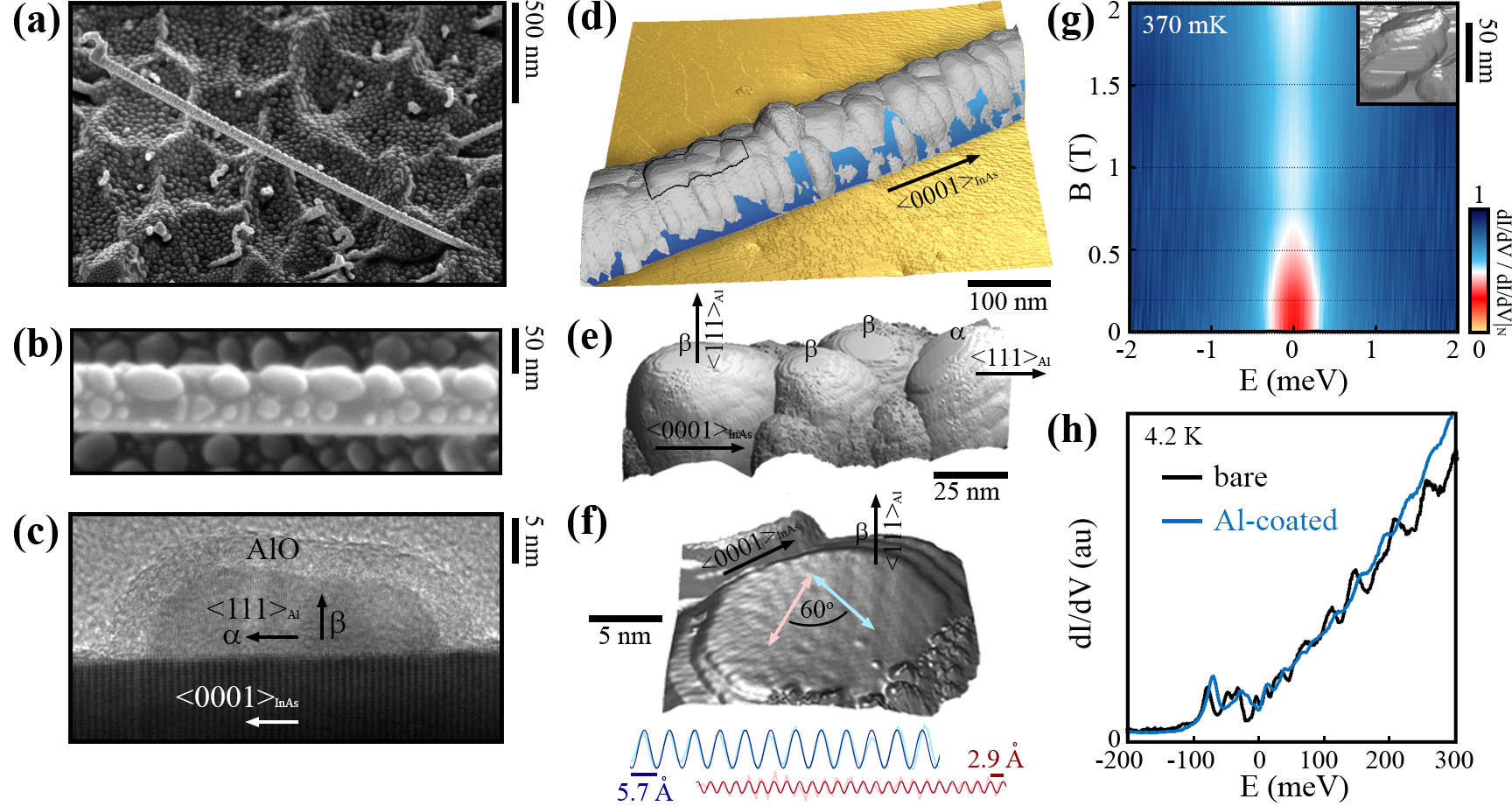} 
\centering
\caption{\label{fig1} \textbf{(a)} SEM image of an InAs nanowire covered with epitaxially grown Al islands. \textbf{(b)} Enlarged SEM image of such a nanowire. The large islands grow on the facet directed towards the Al evaporation cell. They have a slightly elongated shape since their growth is constrained by the 30 nm width of the nanowire facet. Smaller islands form on the side facets. \textbf{(c)} TEM image showing the clean interface between the nanowire and an oxide layer of about 5 nm that builds upon exposure to ambient conditions. The imaged island is composed of two Al crystallites in different orientation of their $<$111$>$ FCC lattice. \textbf{(d)} False color STM topographic image of an InAs nanowire (blue) on a gold substrate (yellow) covered with Al islands (gray). \textbf{(e)} Enlarged topography of the four Al islands marked in (d) showing distinct orientations of the crystallites' top facets corresponding to the $\alpha$ and $\beta$ Al growth directions. \textbf{(f)} The topmost facet of each crystallite is atomically flat, revealing the Al crystal structure. Two topographic profiles taken across that island at the locations marked by pink and light blue arrows are shown in the bottom of the panel in corresponding colors. They are fitted with sinusoidal profiles in red and blue, respectively, revealing a doubled periodicity along one. \textbf{(g)} Interpolated false color plot of a $dI/dV$ measurement on a larger island (topography in inset) showing the closure of the superconducting gap and the Coulomb blockade gap that remains at higher magnetic fields. Dashed black lines correspond to the actual spectra acquired at different magnetic fields and are shown in Fig. \ref{figS0}. \textbf{(h)} Representative $dI/dV$ curves measured on a bare nanowire (black) and an exposed InAs segment of a nanowire deposited with Al islands (blue). In both cases the conduction band onsets at about 100 meV below $E_F$.}     
\end{figure}

We conduct spectroscopic STM study of InAs nanowires coupled to epitaxial Al islands deposited \textit{in-situ} by molecular beam epitaxy (MBE). Reclined Wurtzite InAs nanowires with a typical diameter of about 70 nm, as the one shown by scanning electron microscopy (SEM) in Fig. \ref{fig1}(a), were grown in the $<$111$>$ direction on an epi-ready (001) InAs substrate \cite{Kang2017a}. The nanowires have an hexagonal cross section obtained by low temperature growth procedure subsequent to the growth of the nanowires \cite{Reiner2017HotNanowires}. Epitaxial deposition of Al was carried out \textit{in-situ} at room-temperature. The high diffusivity of the Al atoms on the side facets results in segregation to separate sub-micron islands. The Al islands on the facet that faces the Al cell are a few tens of nanometers long (Fig. \ref{fig1}(b)). Here we also study the Al islands on the two adjacent facets on either side of the hexagonal nanowire, that are much smaller in volume because of the lower flux that impinges on them. An \textit{ex-situ} transmission electron microscope (TEM) image of a single Al island (Fig. \ref{fig1}(c)) shows the atomically sharp interface between the InAs nanowire and the Al island. The island's typical thickness is 5-10 nm. Atomic resolution reveals that the particular island imaged in TEM is composed of an $\alpha$ and $\beta$ crystallites in which the $<$111$>$ direction of the face centered cubic (FCC) Al lattice is oriented parallel and perpendicular to the nanowire $<$0001$>$ axis, respectively \cite{Krogstrup2015EpitaxyNanowires, Kang2017a}. For STM measurements the nanowires were mechanically harvested immediately after their growth and deposited over a clean gold crystal, prior to their transfer to a commercial Unisoku STM. The entire harvest and transfer process was carried out under ultra-high vacuum conditions, in a home-built vacuum suitcase as was demonstrated recently with bare InAs nanowires \cite{Reiner2017HotNanowires}. 

A topographic STM image of Al islands on an InAs nanowire that resides on a gold substrate is shown in Fig. \ref{fig1}(d). Although the islands are physically separated from one another, as can be inferred from the SEM image, in the STM image they appear enlarged and overlapping. This distortion results from the convolution of  the extruded topographic profile of the islands with the microscopic shape of the tip apex. Nevertheless, the imaged top surfaces of the islands, which are scanned by the very end of the tip, are not affected by such distortion. Hence, they  represent the actual structure of the islands top facets. We find that the Al islands terminate by an atomically flat facet. The two facet orientations we find, shown in Fig. \ref{fig1}(e), are indicative of the two crystalline orientations, $\alpha$ and $\beta$, of the Al FCC lattice. After prolonged Al growth these eventually merge into a polycrystalline Al structure. The topographic image of an individual $\beta$ oriented island in Fig. \ref{fig1}(f) shows the ordered crystalline FCC structure of the Al atoms. We present atomically resolved topographic profiles along the two crystallographic directions. Surprisingly, one direction exhibits the expected lattice constant of bulk Al (2.85 \AA), while the other one shows precisely doubled periodicity (5.7 \AA). We attribute the doubling of the imaged unit cell to slight buckling deformation of the Al crystal (of about 1\% of the unit cell) in response to the strain induced by the lattice mismatch between the InAs substrate and the Al island.

\subsection{Fermi level pinning in InAs nanowires}

The atomically pristine Al surfaces allow us to investigate the spectroscopic properties of the hybrid InAs-Al system. We begin by imagining the superconducting state of the Al islands (see also Appendix \ref{App:SC}). The low temperature differential conductance ($dI/dV$) spectrum in Fig. \ref{fig1}(g) was measured on a relatively large Island of about $50$ nm in width. A significant suppression, in the form of a soft gap, appears in the $dI/dV$ around zero bias in the absence of a magnetic field. With application of a magnetic field $B$ perpendicular to the gold substrate, this soft gap gradually narrows until it closes at about $B=700$ mT, in general agreement with transport measurements of mesoscopic Al islands \cite{Grivnin2019}. We identify this low field dip with the superconducting gap of Al, $\Delta_{Al}$. It appears soft due to the measurement temperature, $T=370$ mK, compared to the bulk critical superconducting temperature is $T_C=1.2$ K \cite{Kim2011UniversalStructures, Nishio2006SuperconductivityMicroscopy}. The effective width of the gap at $B = 0$ T is $\Delta \approx 250 \ \upmu \mathrm{eV}$, which is somewhat larger than the bulk value $\Delta_{Al} = 180 \ \upmu \mathrm{eV}$. The extended width of the superconducting gap is contributed by a narrower and shallower gap which is revealed around zero bias at high magnetic fields. The smaller gap appears as a 40\% suppression in the $dI/dV$, and remains unchanged up to the highest field applied of 2 T. In our spectroscopic measurements of different Al islands we have detected similar gaps around zero bias of widths far greater than $\Delta_{Al}$. We identify these with Coulomb blockade gaps induced due to electronic charging effects introduced by the high interface resistance and small capacitance of the mesoscopic Al islands. The Coulomb blockade regime will be discussed in detail below.

We first examine the effect the Al islands have on the semiconducting properties of the InAs nanowire. In Fig. \ref{fig1}(h) we compare a typical $dI/dV$ spectrum measured in between Al islands on an exposed bare InAs segment  with a spectrum we have measured on a completely bare nanowire \cite{Reiner2017HotNanowires} (blue versus black lines, respectively). In both cases we find that the conduction band of InAs is quantized into a series of peaks signifying the Van-Hove singularities occurring at the bottom of each subband in the nanowire. The peaks are fairly equally spaced in energy and ride an approximately linearly increasing density of states, both reminiscent of the approximate linear dispersion of the conduction band in InAs. Remarkably, the onset of the conduction band above the semiconducting gap occurs in both cases about 100 meV below the Fermi energy $E_F$ (zero bias in STM). We stress that more than 70$\%$ of the top facet of the Al deposited nanowire is covered by epitaxial Al islands (Fig. \ref{fig1}(d)). This is a direct indication that epitaxial Al coating, which was shown to be necessary to induce a hard superconducting gap in the nanowire \cite{Chang2015HardNanowires}, does not lead to any substantial charge doping of the InAs nanowires.

\begin{figure}
\includegraphics[width=0.6\linewidth]{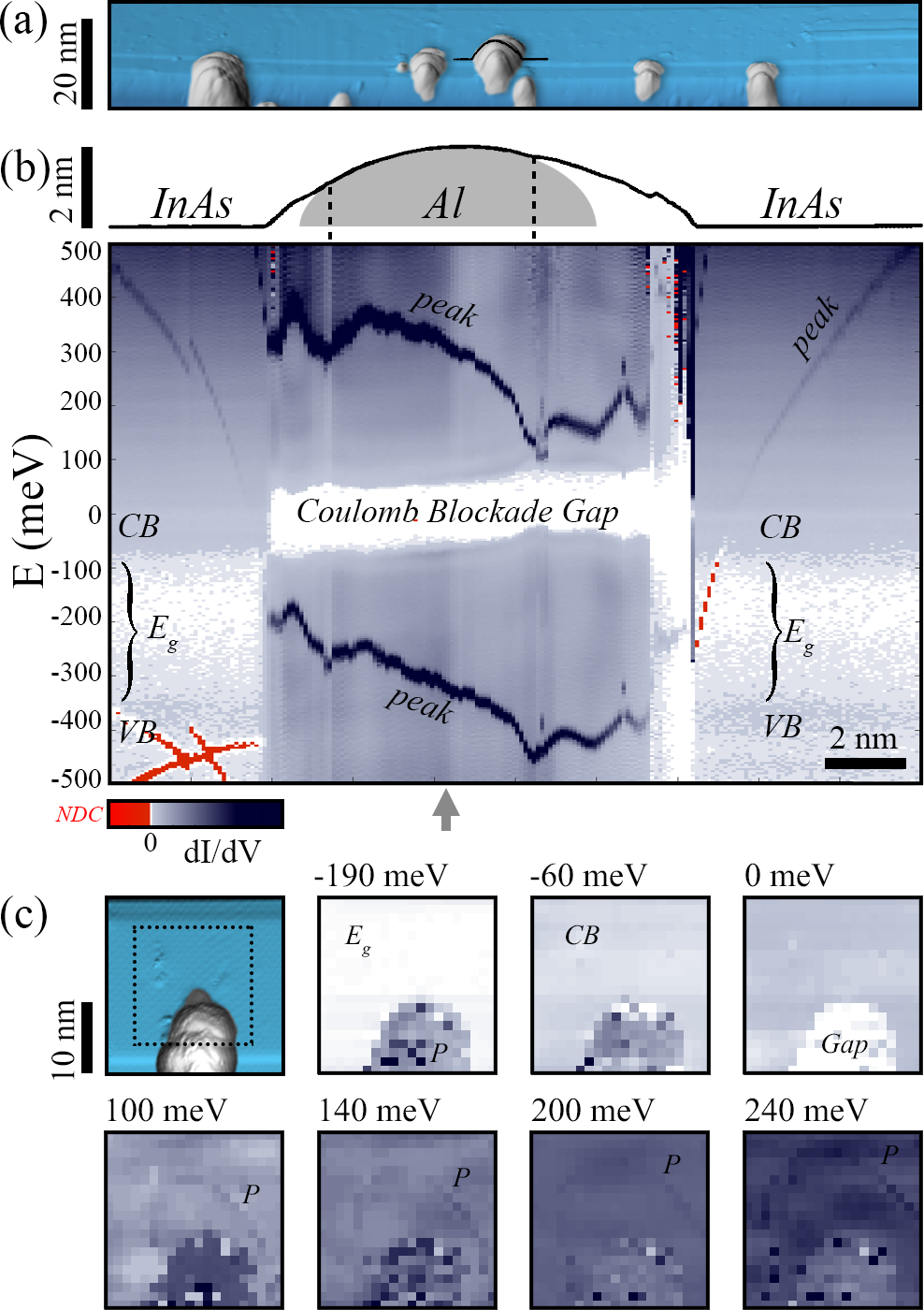} 
\caption{\label{fig2} \textbf{(a)}  STM topography of a side facet with small aluminum droplets. The black line relates to the topographic profile of the island we study in (b). The lateral dimensions of the island are slightly smeared by the blunt apex of the tip, as evident from deviations from the smooth curvature on both sides of the nanowire concurrently with sharp changes in the trend of the $dI/dV$, but they can be estimated to be about 8 nm across and 2 nm in height. \textbf{(b)} A spectroscopic linecut corresponding to the topographic profile  across the Al island and the InAs nanowire adjacent to it. Around $E_F$ the Coulomb blockade regime can be seen, along with dark-blue resonances in the electron and hole sectors. These resonances leak into the nanowire and sharply disperse as a function of position. The resonances change sharply from positive to negative differential conductance (NDC) (marked in red) across the onset of the conduction band (above $ -100 \ \mathrm{meV}$). \textbf{(c)} Conductance maps of a different small island showing the spatial dependence of the different gaps and resonances on the Al island and the InAs nanowire.}
\end{figure}

We further investigate the spectral properties of the InAs nanowire at the immediate vicinity of Al islands. STM topography of several small Al islands formed on the nanowire side facet is shown in Fig. \ref{fig2}(a). A topographic profile of one of the islands and a $dI/dV$ linecut taken across it are shown in Fig. \ref{fig2}(b). The $dI/dV$ measured on the bare InAs segments on either side of the Al island, indicates that the onset of the InAs conduction band above the semiconducting gap ($E_g$) remains unchanged regardless of how close to the Al island we probe it. In particular, we do not detect any traces of band bending directly associated with the presence of the metallic island. We find similar behavior also at the epitaxial interface of the InAs nanowire with the Au droplet at its end that catalyzed the nanowire growth, discussed in Appendix \ref{App:gold}. This further asserts that the robust pinning of the Fermi-level in InAs nanowires does not result from its interfacing with a particular metallic contact, but is rather a general property of those semiconducting nanowires. Such a robust behavior was not identified in recent Schr\"{o}dinger-Poisson models that typically predict a substantial band bending of several tens of meV that will be induced by interfacing InAs nanowires with Al \cite{Woods2018EffectiveDevices, Antipov2018EffectsNanowires, Mikkelsen2018HybridizationInterfaces}.  The screening of this effect (usually calculated normal to the interface) is expected to occur over a scale of several tens of nanometers due to the low carrier density of the semiconductor. A naive calculation of the Thomas-Fermi level screening length that accounts for the Fermi energy lying 100 meV above the bottom of the conduction band also estimates it to be longer than 10 nm. Accordingly, our observations put a stringent constraint on the occurrence of metal induced band bending in InAs nanowires suggesting that $E_F$ is pinned to its bare nanowire value.

Fermi level pinning, which refers to the insusceptibility of the bulk Fermi level to surface perturbations, is known to occur in semiconductors whenever they host metallic surface states \cite{monch1995semiconductor, Bardeen1947SurfaceStates}. Since the populated surface states are electrically charged they are accompanied by a spatially extended charge layer of opposite polarity which screens the induced surface charge. As a result, a charge dipole forms at the surface of the semiconductor. We now investigate the effect of such surface dipole on the resistive interface between the InAs nanowire and the Al islands deposited on its facets. For this purpose we examine the electronic conduction properties of the Al Islands in the Coulomb blockade regime. In our setup, this occurs when the tip-Al and InAs-Al tunnel resistances both surpass the quantum of resistance, $R_{Q} \approx 13 \ \mathrm{k \Omega}$, and the charging energy $E_C = e^2/2C$ of the island ($C$ being the island’s total geometrical capacitance) is greater than the thermal energy, $E_C \gg k_B T$.  The smaller the Al island is, the smaller its capacitance and the larger is its charging energy. The interface resistance between the nanowire and the Al island, $R_{NA}$, also grows as the island shrinks in size (the tip-Al tunneling resistance is set to 1 G$ \Omega$ throughout this study). Accordingly, in smaller islands the Coulomb blockade phenomenology is more pronounced. 

In presence of nano-scale Al islands a variety of spectral features appear, as seen in Fig. \ref{fig2}(b). They entail a wealth of information on the hybrid InAs-Al system. A hard gap of about 120 meV appears around zero bias when tunneling directly into the Al island. At energies beyond this gap the $dI/dV$ gradually attains a constant value on top of which sharp peaks appear at both the electron and hole sectors. The peaks shift strongly in energy as a function of location across the island while retaining their energy separation constant. The peaks extend into the bare nanowire segments adjacent to the Al islands. Here, however, they shift monotonically to higher biases as a function of the distance from the island. At energies below the semiconducting gap (i.e. at $E<-100 \ \mathrm{meV}$) these peaks show negative differential conductance (highlighted in red) \cite{Muzychenko2009}. We have detected similar phenomena at the vicinity of the polycrystalline gold droplet at the end of the InAs nanowire that catalyzed its growth (see Appendix \ref{App:gold} for more details). 

The full spatial structure of these spectral features can be appreciated from the constant-energy $dI/dV$ maps taken on another, slightly larger (about 10 nm across), Al island shown in Fig. \ref{fig2}(c). At low energies (-190 meV), the finite $dI/dV$ of the metallic island (light blue), including spatially scattered conductance peaks (dark blue pixels), contrasts with the vanishing $dI/dV$ of the semiconducting gap of the InAs nanowire that surrounds the island (white). This contrast reduces above the onset of the conduction band (shown at -60 meV), and then reverses closer to zero bias, where the Al island becomes gapped. Al induced band bending would have appeared as a dark halo of increased $dI/dV$ surrounding the island at these energies, which is clearly absent in the spectral images. At yet higher energies conductance peaks are scattered at different positions inside the island, while a sharp ring-like structure (marked with $P$) surrounds the island and disperses outwards with increasing bias.        

\begin{figure}
\includegraphics[width=0.5\linewidth]{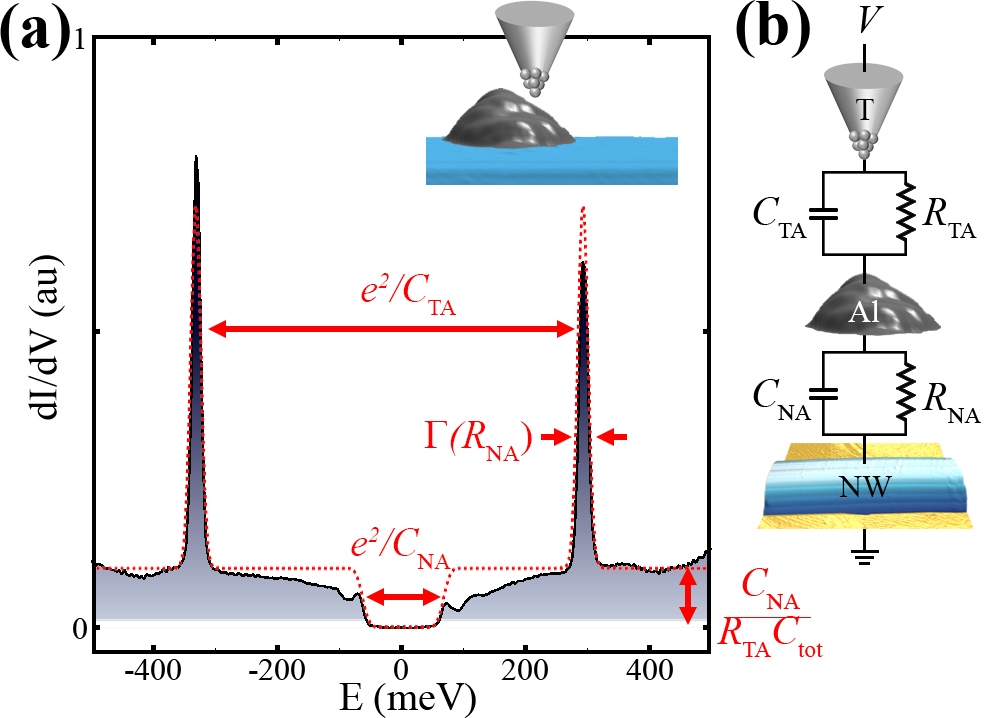}
\centering
\caption{\label{fig3} (a) $dI/dV$ at a particular position on the droplet (inset, indicated by arrow in Fig. \ref{fig2}(b)), with a fit to the simplified DBTJ model (red dashed line). The annotations refer to the circuit parameters that can be extracted from the measurement. (b) Schematic diagram of the equivalent circuit when the STM tip is positioned on the droplet.}
\end{figure}

We now focus on the spectral features seen on the Al islands. An individual $dI/dV$ curve, at a position marked with an arrow in Fig. \ref{fig2}(b), is displayed in Fig. \ref{fig3}(a). The detection of a Coulomb blockade gap and conductance peaks is typical to previously studied tunnel coupled quantum dots, asserting that a tunneling barrier forms not only across the tip-Al junction, but also at the InAs-Al interface. Accordingly, the tip-Al-nanowire system can be considered as a double barrier tunneling junction (DBTJ) where each of the two junctions is characterized by a resistive and a capacitive coupling. The equivalent circuit is outlined schematically in Fig. \ref{fig3}(b). It accounts for capacitive and resistive couplings between the STM tip and Al island as well as between the Al island and the InAs nanowire. Although the resistive tunneling barriers are highly non-linear, the DBTJ problem can be solved analytically \cite{Amman1991}. The solution is greatly simplified whenever one of the tunneling junctions is much more resistive than the other \cite{Hanna1991}. In our setup the nanowire-Al tunneling resistance is thought to be negligible relative to the tip-Al junction, maintained at $R_{TA}=$1 G$\Omega$. In this limit the smaller resistor, $R_{NA}$, sets the most probable number of electrons occupying the island, $n_0$, while the current flow through it is regulated by the larger resistor, $R_{TA}$. The semi-classical solution is then given by:
\begin{equation}
\label{eq2}
  I(V)=
\begin{cases}
0, & -e/2<C_{NA}V+Q_0<e/2 \cr
\frac{Q_0 - n_0(V) e +C_{NA}V - \operatorname{sgn} (V) e/2}{R_{TA} (C_{TA}+C_{NA})}, & \mathrm{otherwise} \\
\end{cases}
\end{equation}

 where $C_{NA}$ and $C_{TA}$ are the nanowire-Al and tip-Al mutual capacitances, respectively, and $Q_0$ is a polarization charge induced on the island in presence of its electrostatic environment. The number of electrons accumulated on the island, $n_0$, is an integer that satisfies:
 \begin{equation}
 \label{eq3}
 -e/2 < en_0 + C_{TA}V - Q_0 < e/2.    
 \end{equation}
  Here we have further assumed that the tunneling rates are independent of the density of states $\nu(E)$ of both the island and the nanowire, and that the transmission through the tunneling barrier, $T(E)$, is energy independent. We will revisit these assumptions later. Despite those simplifying assumptions our theoretical calculation captures well most of the spectroscopic features, as demonstrated by the fitted dotted line in Fig. \ref{fig3}(a). The voltage range around zero bias showing zero current marks the Coulomb blockade regime, where electrostatic repulsion blocks electron accumulation and transport through the island. From the energy width of the Coulomb blockade gap we extract $C_{NA}=1.5$ aF. Intriguingly, if we model this capacitive coupling with a plate capacitor, $C_{NA} \approx \varepsilon A / d$, and substitute the bulk dielectric of InAs $\varepsilon=15 \varepsilon_0$ and an interface area of $A=80$ nm$^2$ extracted from the island topography, we find the effective electronic layer is at a distance of $d \approx 10$ nm from the InAs-Al interface.

Once the voltage surpasses the charging energy of the island (beyond the voltage range specified in Eq. \ref{eq2}) current flow onsets. The semi-classical approximation predicts a linear I-V relation which is periodically offset by the island occupation number, $n_0$, whenever it becomes energetically favorable to let an additional electron enter, as expressed in Eq. \ref{eq3}. Accordingly, sharp steps are expected in $I(V)$ whenever the voltage bias is large enough to allow an additional electron occupy the island. These steps translate to the peaks we observe in the $dI/dV$ profile as we scan the bias voltage. From the voltage separation between sequential Coulomb peaks we read off $C_{TA}$=0.3 aF. We find that, up to an overall scaling factor, the Coulomb peaks and Coulomb blockade gap exhibit anti-correlated shift in energy as the STM tip traverses the Al island (see Fig. \ref{figS4}(a) in Appendix \ref{App:RE}). Since the former is dominated by the tip-Al capacitance (Eq. \ref{eq3}) and the latter by the nanowire-Al capacitance (Eq. \ref{eq2}), they are oppositely affected by the charge accumulated on the Al island, $Q_0$, that varies with the STM tip position. As the STM tip descends from the Al island to the bare nanowire facet the resistive tunnelling channel between them, $R_{TA}$, is cut off exponentially, but the capacitive coupling, $C_{TA}$, remains finite as it decays as a power law with distance. This alternative setting, that gives rise to a negative differential conductance signal, is analyzed in Appendix \ref{App:Off}, and demonstrates the ability of the model to describe the DBTJ nanowire-island-tip system we measure. The capacitive coupling may also support future charge sensing of superconducting island occupation, desirable for Majorana detection \cite{Ben-Shach2015DetectingMajorana}.

An intriguing system parameter we investigate is the resistivity of the buried InAs-Al interface. The observation of Coulomb blockade gaps and peaks on the Al islands signifies the existence of a resistive barrier at their interface with the nanowire. To estimate it quantitatively we note that the energy width of the $dI/dV$ peak in Fig. \ref{fig3}(a) is larger than the instrumental broadening (temperature broadening is about $1 \ \mathrm{meV}$ and the AC excitation amplitude is 5 meV). 
The resistance of this barrier can be extracted from a fit to the DBTJ semi-classical model (see Appendix  \ref{App:Orthodox}), yielding $R_{NA}=1.3  \pm 0.3 \ \mathrm{M}\Omega$. For the corresponding interface area, $A \approx 100$ nm$^2$, this yields interface resistivity of $\rho_{NA} \approx 1.3 \ \upmu \Omega \ \mathrm{cm}^2$ (in Appendix \ref{App:RE} we substantiate that such a simple scaling is valid even for relatively small contact area). Such resistivity is generally considered as a good metal-semiconductor contact. Ultimately, such hybrid InAs-Al devices are aimed towards inducing topological superconductivity in one-dimensional structures by the combination of high spin-orbit coupling, Zeeman field and proximity induced superconductivity. While the chemical potential should remain fairly constant along the superconducting segment, which seems to be guaranteed by Fermi level pinning, the proximity effect requires strong hybridization of the electronic wavefunctions in the nanowire and in the Al island. It is desirable that the interface resistance would be lower than a quantum of resistance $R_Q=h/2e^2 \approx 13$ K$\Omega$ \cite{Hammer2007DensityScattering}. Considering the resistivity value we measured we conclude that even short Al segments, of about 200 nm (assuming typical nanowire facet width of 50 nm) are sufficient to induce superconducting correlations in the nanowire. However, shorter Al coverage will lead to a strong suppression of the proximity effect. These experimental values should be taken into considerations when designing Majorana quantum dot devices where the charging energy is harnessed to prevent thermally excited quasi-particle from entering the topological device \cite{Karzig2017ScalableModes, Plugge2016RoadmapCodes}.

\subsection{Energy barrier at the InAs-Al interface}

\begin{figure}
\includegraphics[width=0.5\linewidth]{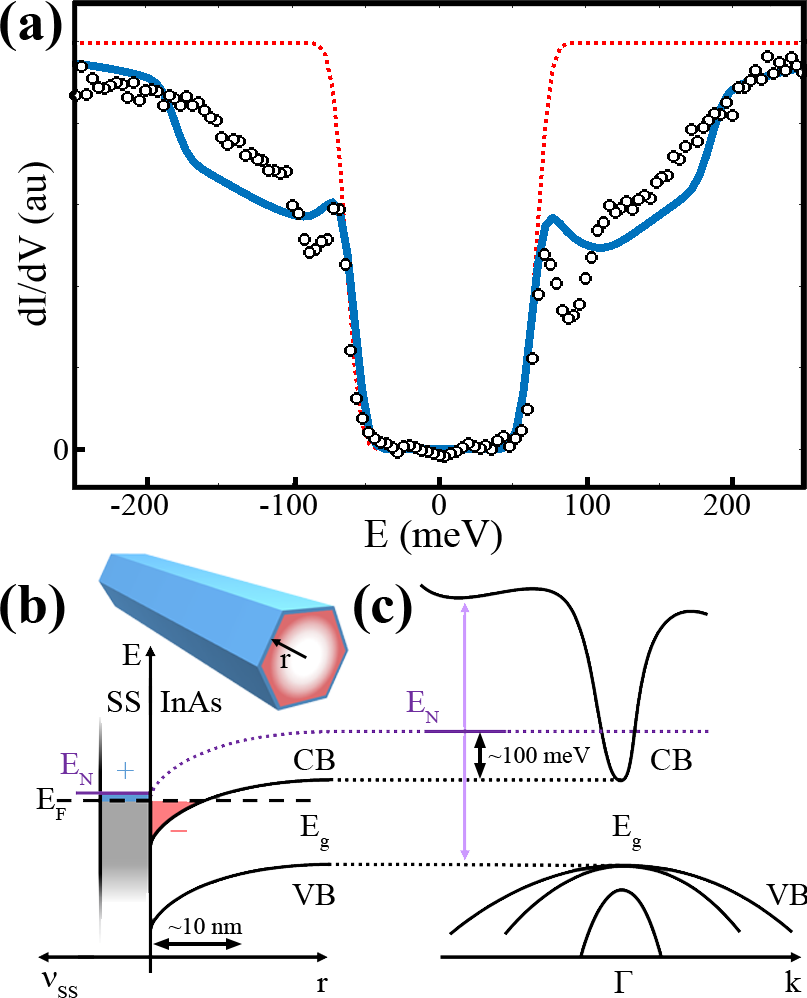}
\centering
\caption{\label{fig4} Low energy $dI/dV$ signal (circles) fitted with the simplified model (dotted line) and the model that accounts for the energy-dependent transmission through the interface, $T_{NA}(E)$ (solid line). (b) Schematic band diagram illustrating the positively charged surface states versus negatively charged accumulation layer. Inset show schematically the radial charge distributions at the nanowire surface. (c) InAs band structure in which the neutrality point of the surface states, $E_N$, resides in the middle of the indirect gap, and thus lies 100 meV into the conduction band.}
\end{figure}

We now address the main discrepancy between our simplified model and the measurement. While the model predicts a sharp recovery of the $dI/dV$ signal beyond the Coulomb blockade gap, demonstrated by the dotted line in Fig. \ref{fig4}(a), the measured $dI/dV$ recovers gradually and displays additional fainter peak structure right at its onset. The emergent peak decorates the rims of the Coulomb blockade gap all across the island. We note that a sharp Coulomb peak is expected to appear at the edge of the Coulomb blockade gap in the opposite regime, where the InAs-Al interface resistance approaches or exceeds that of the the tip-Al tunneling junction, $R_{TA} \leq R_{NA}$, while maintaining the same capacitance ratio, $C_{TA}<C_{NA}$. We therefore reexamine our approximation for the InAs-Al tunneling rate by considering its full semiclassical expression that allows for comparable tunneling resistances due to energy dependent tunneling barriers (Appendix \ref{App:Orthodox}) 
\begin{equation}
\label{eq5}
\Gamma_{NA}(N,\Delta E_{NA})=\Theta(\Delta E_{NA})\nu_N(0)\nu_A(0)\int_{-\Delta E_{NA}}^0 |T(E)|^2dE,
\end{equation}
where $\Theta$ is the Heaviside function, $\Delta E_{NA}=\varepsilon_F^N-\varepsilon_F^A$ is the voltage drop across the nanowire-Al junction, $\varepsilon_F^N$ and $\varepsilon_F^A$ are the chemical potentials on the nanowire and Al island, respectively, and $\nu_{N}$ and $\nu_{A}$ are the density of states of the nanowire and Al island, respectively, which are still assumed to be energy independent since both have a rather smooth behavior at those energies.  

We thus allow for an energy dependent transmission $T_{NA}(E)$ through the interface. For the transmission function we consider electron tunneling through a square potential barrier with an atomic scale width of $w=1$ nm, and an adjustable energy scale $V_{0}$. It yields $T(E)=[1+V_0^2 \sinh^2(kw)/4E(V_0-E)]^{-1/2}$, where $k=\sqrt{2m^*(V_0-E)/\hbar^2}$, $m^*=0.05m_e$ is the effective mass of the electron in InAs, and $m_e$ is the bare mass of the electron. Substituting this into Eq. \ref{eq5} with $V_0=$0.5 eV reproduces the observed additional low energy features, seen in Fig. \ref{fig4}(a), of a peak at the edge of the Coulomb blockade gap followed by a gradual recovery of the $dI/dV$ signal which is set by a combination of $R_{NA}(V)$ and $R_{TA}$. As the bias increases above the potential barrier, $V > V_0$,  the resistance ratio crosses back to $R_{NA} \ll R_{TA}$. We thus conclude that the appearance of the peak at the edge of the Coulomb blockade gap signifies a finite potential barrier that the electrons need to surmount across the InAs-Al interface. Here, we have analyzed the two extreme cases of low and high bias separately. A more comprehensive model that would account for the relative role of the capacitances at the intermediate regime where the resistances are of the same order is needed.

\subsection{Surface states in Semiconducting nanowires}

Our observation of Fermi-level pinning, including the exact energy value it is pinned at, alludes to the existence of surface states that are tightly bound to the facets of the semiconducting nanowires \cite{Halpern2012}, as illustrated in Fig. \ref{fig4}(b). Theoretically, these are accommodated within the Bardeen model \cite{Bardeen1947SurfaceStates}, whose implementation to semiconducting nanowires we discuss here. Unlike the Schottky-Mott model that analyzes semiconductor-metal interfaces \cite{Schottky1939, Mott1939}, the Bardeen model further accounts for the presence of surface states preexisting on the nanowire facets. Previous reports suggest the formation of such states on surfaces of bulk semiconducting samples is avoided only by an atomically flat cleave of a non-polar surface of a single crystal under ultra-high vacuum \cite{Vlaic2017a, Zhang2018QuantumInAs}. Though the \{11-20\} side facets of our $\langle0001\rangle$ InAs nanowires are non-polar, and in spite of their growth, transfer and measurement under ultra-high vacuum conditions, their complex growth protocol and imaged surface purity are not compatible with those stringent conditions. In ref. \cite{Reiner2017HotNanowires} we imaged in topography  both crystallographic defects as step edges and stacking faults, as well as atomic surface impurities in a concentration which is sufficient to induce metallic surfaces even before metallization of the surface by superconducting elements. We find the chemical potential to lie about 100 meV above the bottom of the conduction band both in bare nanowires as well as in nanowires deposited with Al. This value corresponds to a surface electron density of about $10^{12} \ \mathrm{cm}^{-2}$, which is just about the value at which Fermi-level pinning is known to set in \cite{monch1995semiconductor,luth2010solid}. At such densities the surface states become so dense  that surface modification, as deposition of metals, result in negligible effect on the electronic state of the semiconducting nanowire. 

The pinning value of the Fermi level in the presence of occupied surface states is set by the energy of their neutrality point, $E_N$, which separates donor-like states below it to acceptor-like states above. The change in their nature is inherited from the bulk band structure and occurs approximately at the center of the semiconducting gap. However, in narrow gap semiconductors, as InAs and InSb, which attract vast attention due to their strong spin-orbit couplings, the neutrality point occurs at the center of the 1 eV indirect gap to the high-density shallow electron pocket at the $L$ point, rather than the 350 meV direct gap to the low-density narrow pocket at $\Gamma$ (Fig. \ref{fig4}(c)) \cite{monch1995semiconductor, luth2010solid}. Consequently, when the bulk Fermi level resides within the semiconducting gap, non-equilibrium occupation is induced on the positively charged surface. In response, this surface potential bends the bands at the vicinity of the surface as to lower the neutrality point towards the Fermi level and discharge the surface states. 

While tuning the neutrality point, the band bending induces a negatively charged accumulation layer over the scale of the semiconductor’s screening length, depicted in Fig. \ref{fig4}(b). Equilibrium is achieved by charge neutrality when the densities of surface and subsurface opposite charges equate. Since the density of states of the surface states is much higher than that of the semiconductor, the surface equilibrates by bringing the Fermi energy to close vicinity of the neutrality point. This peculiar dominance of the indirect gap pins the Fermi-level in InAs bulk crystals and, as we show here, in nanowires at about 100 meV within the conduction band. The equilibrium charge distribution of positively charged surface states backed by a negatively charge accumulation layer thus forms a radial surface dipole normal to the nanowire facets. We attribute the potential barrier that leads to the energy dependent transmission through the nanowire interface to that surface dipole. Its presence may very well dictate the ability to induce superconducting correlation into the core states of the nanowire. On the other hand, the surface dipole may very well give rise to a strong Rashba-type spin-orbit coupling which is necessary for inducing topological superconductivity.

The presence of surface states with high density, which is ubiquitous to III-V semiconductors but was overlooked in recent modellings of nanowire-metal interfaces \cite{Woods2018EffectiveDevices, Antipov2018EffectsNanowires, Mikkelsen2018HybridizationInterfaces}, relieves the susceptibility of the semiconducting nanowire to various surface perturbations (including invasiveness of the STM tip) and specifically to the introduction of metallic islands \cite{dassarma2017dot}. In turn, this puts in question the susceptibility of the electronic states in the nanowire to Fermi-level tuning by gating. Our preliminary results, presented in Appendix \ref{App:gating}, provide first indication for a rectified behavior in bare InAs nanowires. We find that positive gate voltages do not lead to upward shift of the chemical potential due to its pinning. However, sufficiently large negative gate voltages allow its depinning and consequently a downward shift of the chemical potential towards the semiconducting gap that corresponds to a depletion of the nanowire from its charge carriers. More studies are needed to fully characterize this response in bare and in metal deposited nanowires.

\section{Summary}
In summary, we have investigated superconducting Al islands epitaxially grown \textit{in-situ} on InAs nanowires. In spite of the direct interface between the two materials, global charge transfer as well as excess metal-induced local band bending are both absent from the measured local $dI/dV$ spectra. The charging energy of the Al islands and a resistive barrier at their interface with the nanowire establish Coulomb blockade dynamics for the tunneling electrons consisting of a gap and peaks. By studying those we characterize the energy profile of the resistive barrier at the buried interface. Beyond the interface resistance we extract at high energies, we find a potential barrier that further suppresses transmission of electrons at low bias. While constraining wavefunction hybridization which is desired for inducing superconducting correlations from the islands to the nanowire, our measurements and modelling demonstrate the robustness of the electronic properties of the semiconducting nanowire to surface perturbations and metal deposition. They offer new means to probe and manipulate Majorana end modes in such hybrid devices.

\begin{acknowledgments}
HB acknowledges funding by the Israeli Science Foundation (ISF), the Minerva Foundation, and the European Research Council (ERC, project ‘TOPONW’). HS acknowledges partial funding by Israeli Science Foundation (grant No. 532/12 and grant No. 3-6799), BSF grant No. 2014098 and IMOS-Tashtiot grant No. 0321-4801. YO Acknowledges funding by the Israeli Science Foundation (ISF), the Deutsche
Forschungsgemeinschaft (CRC 183), the Binational Science Foundation (BSF), the European Union's Horizon 2020
research ERC and innovation programme (grant agreement 
LEGOTOP No 788715). YO acknowledge partial support by the European Union’s Horizon 2020
research and innovation programme (grant agreement
LEGOTOP No 788715), the DFG (CRC/Transregio 183,
EI 519/7- 1), and the Israel Science Foundation (ISF) and
the Binational Science Foundation (BSF).
\end{acknowledgments}

\appendix

\section{Superconductivity of the Al Islands}
\label{App:SC}

\begin{figure}
\includegraphics[width=0.4\linewidth]{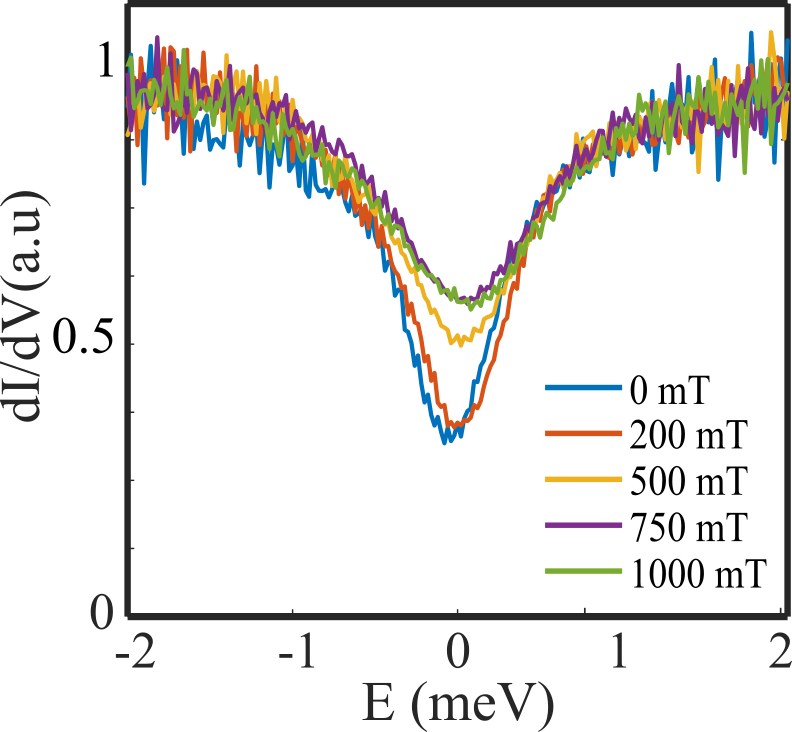} 
\centering
\caption{\label{figS0} $dI/dV$ measurement at 370 mK of the magnetic field dependence of the spectra, taken on a larger island. The dip diminishes with B, confirming it originates from the superconductivity. The interpolated spectra and the island's topography are shown in Fig. \ref{fig1}(g).}    
\end{figure}

To investigate the spectroscopic properties of the superconducting state of the electrons in the Al islands we show a low temperature measurement ($T=370$ mK) of one of the largest islands we scanned (Fig. \ref{figS0}). A significant dip of almost 70\% appears in the $dI/dV$ at zero bias in the absence of a magnetic field. With application of a magnetic field $B$, the spectrum evolution shows a hierarchy of two zero bias gaps. One gap is susceptible to magnetic field and the other is insusceptible, and is identified as the Coulomb blockade gap as elaborated in the main text. The low-field gap gradually decreases with $B$ applied perpendicular to the gold substrate and closes at about $B=700$ mT, in general agreement with similar nanowire tunneling devices studied by transport measurements \cite{Grivnin2019}. We thus identify this dip in the $dI/dV$ with the superconducting gap of Al, and attribute its soft appearance to the measurement temperature that is comparable to the bulk critical temperature $T_C=1.2$ K (\cite{Kim2011UniversalStructures, Nishio2006SuperconductivityMicroscopy}. The effective width of the gap at $B = 0$ T is $\Delta \approx 250 \ \upmu \mathrm{eV}$, which is about $50 \%$ larger than the bulk value $\Delta_{Al} = 180 \ \upmu \mathrm{eV}$. The increased width of the superconducting gap is contributed by a narrower and shallower gap which is revealed at high magnetic fields, as well as by an excess voltage drop over the interface between the island and the nanowire.  

\section{Fermi level pinning on different crystal termination}
\label{App:pinning}

\begin{figure}
\includegraphics[width=0.4\linewidth]{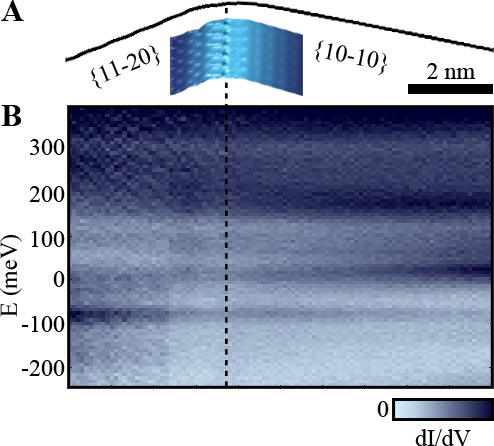} 
\centering
\caption{\label{figS5} $dI/dV$ linecut taken across the \{11-20\} and \{10-10\} side facets. The onset of the conduction band can be seen slightly above -100 meV, and does not vary as a function of position within each facet and between facets.}    
\end{figure}

Fig. \ref{figS5} shows a spectroscopic linecut across both \{11-20\} and \{10-10\} Wurtzite side facets (topography shown in the inset), corresponding to {110} and {112} Zincblende facets. This linecut demonstrates that the onset of the conduction band is fixed across both facets. We therefore conclude that the surface states leading to Fermi-level pinning exist on both facets. The slight intensity evolution away from the interface between the two facets may very well be due to the wavefunction distribution of the quantized electrons within the nanowire geometry.  

\section{Coulomb blockade regime at the poly-crystalline gold catalyst at the nanowire end}
\label{App:gold}

\begin{figure}
\includegraphics[width=\linewidth]{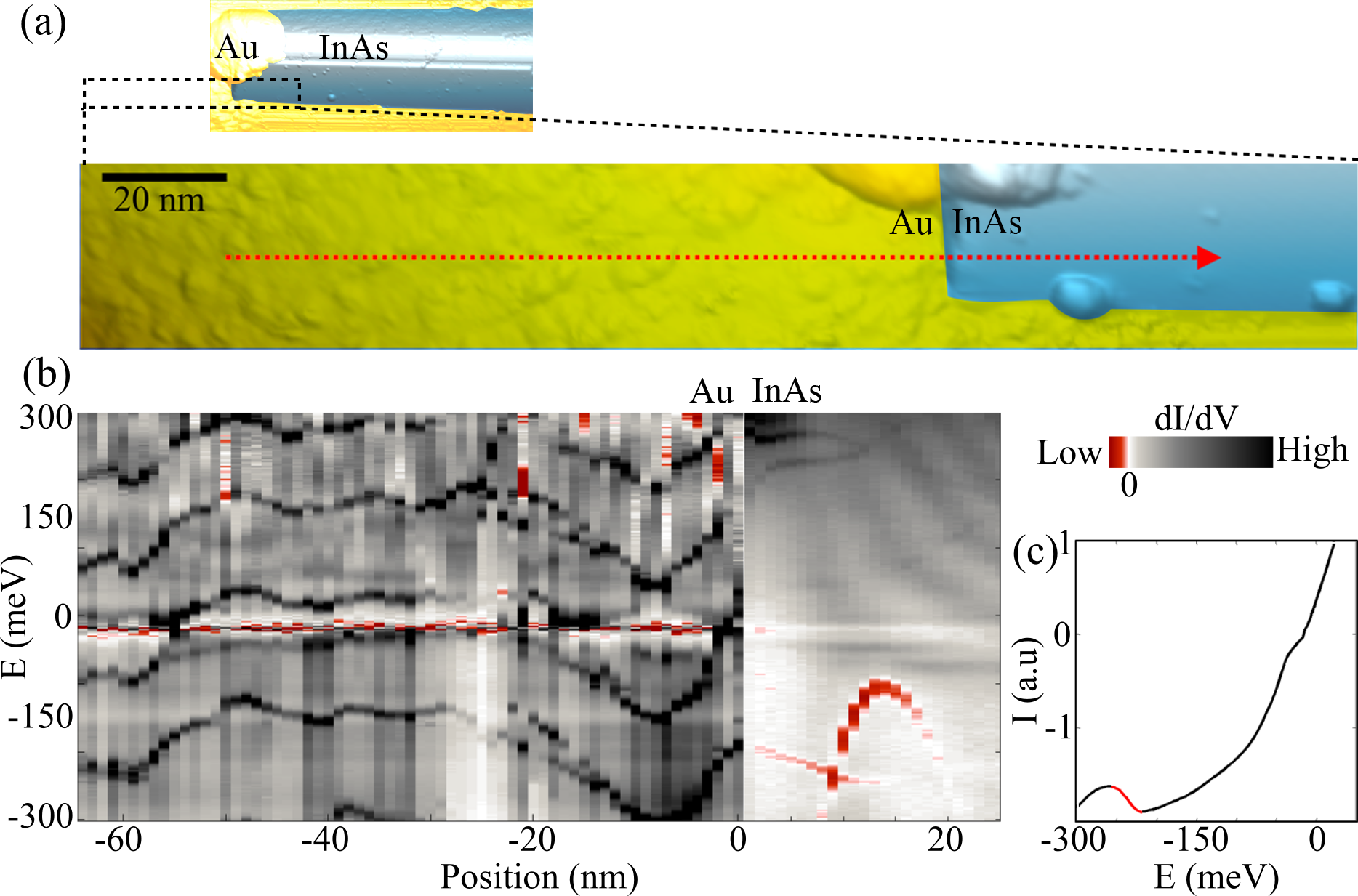} 
\centering
\caption{\label{figS2} \textbf{(a)} Topographic images of the interface between an InAs nanowire and the gold catalyst. The disordered surface of the gold particle may indicate the granularity or poly-crystalline structure that gives rise to quantization. \textbf{(b)} $dI/dV$ map along the red dashed arrow in (a). The spectrum on the gold particle has distinct resonances, with charge disorder that induces fluctuations in the energy of the resonances. The resonances leak into the nanowire where they coexist with the 1D QPI pattern. Within the semiconducting gap ($ E < -75 \ \mathrm{meV}$) the resonances feature negative differential conductance, saturated to red in the false color scale. \textbf{(c)} The integrated current at a position featuring negative differential conductance resonance, where the current is carried by the 1D modes of the nanowire . The red part illustrates decreasing (negative) current when increasing (negative) bias voltage, indicting current blockage of these modes.}  
\end{figure}

We present a different realization of an interface between a nanowire and a mesoscopic metallic system, that shows very similar spectroscopic features. The topography of the interface between a nanowire and the Au particle that catalyzes the growth is shown in Fig. \ref{figS2}(a). In Fig. \ref{figS2}(b) it can be seen that the conductance measured on the gold particle shows similar resonances to those seen on the Al droplet. The absence of a distinct Coulomb blockade gap may indicate that the capacitance of the effective Coulomb island to its environment is larger in this case resulting in small $E_C$. A Coulomb staircase structure on a granular gold film has been reported previously \cite{Hanna1991}, and we assume similar granularity in the disordered polycrystalline gold catalyst is what gives rise to this effective realization of a double barrier tunnel junction, with grain sizes corresponding to a couple of nm. On the InAs, in coexistence with qusiparticle interference patterns reported in ref. \cite{Reiner2017HotNanowires}, we observe the tails of the resonances that extend into the nanowire, and feature position-dependent energy dispersion. The energy dispersion is more complicated in this case, preventing us from analyzing them in a similar fashion to Fig.   \ref{figS1}(c). This is probably because the complex morphology and electrostatic environment renders the capacitive coupling to the tip non-trivial. Remarkably, two of these resonances, those that inhabit the semiconducting gap, are in fact negative. Fig. \ref{figS2}(c) shows the integrated current at negative voltage bias, where the negative differential conductance is manifested by a decrease in the total current supported by the nanowire's 1D modes. The absence of a direct tunneling between the tip and the island, indicates that a variation in the charge distribution due to gating from the tip is responsible to blocking some of the current. 

\section{Energy Dependence of the Coulomb Peak}
\label{App:RE}

\begin{figure}
\includegraphics[width=0.4\linewidth]{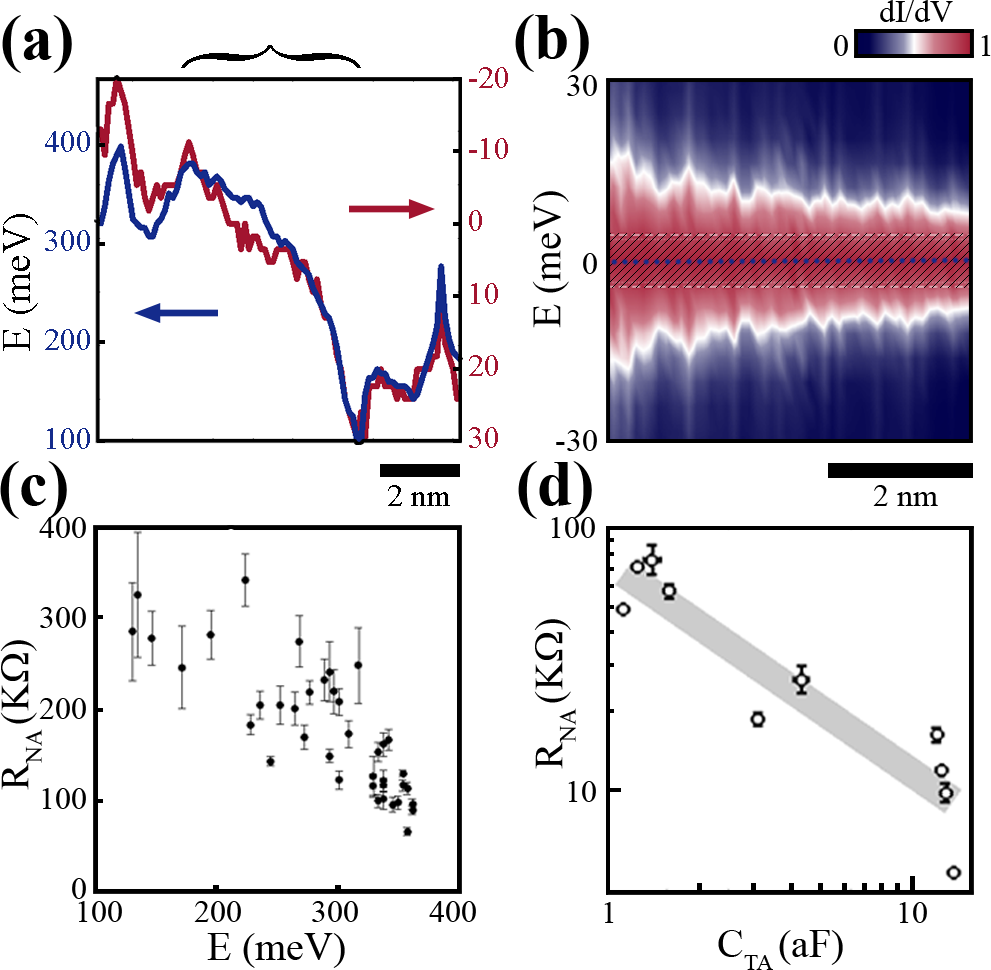} 
\centering
\caption{\label{figS4} (a) Trace of the resonance peak (Fig. \ref{fig2}(b)) showing its peak position (blue) and the mid-gap position (red), illustrating their anti-correlated dispersion. Note the opposite and scaled corresponding voltage axes. (b) A zoomed-in bias window of the $dI/dV$ centered around the dispersing resonance peak from Fig. \ref{fig2}(b). The shaded region indicates the instrumental broadening (c) A plot of $R_{NA}$ at different bias voltages extracted from the energy broadening shown in (b). It confirms that the tunnel barrier is energy dependent, and that closer to the bottom of the band the resistance increases. (d) Log-log plot indicating the power-law behavior, $R_{NA} \sim C_{NA}^{-\alpha}$, expected from the geometrical dependence of these parameters with $\alpha \approx 0.8$.}    
\end{figure}

The interface resistance extracted in Fig. \ref{fig3}(a) was probed through the width of the Coulomb peak at about 300 meV bias as found at a specific tip location on the Al island. However, we find that the peak shifts in energy as the tip shifts in position across the Al island, and with it varies the peak energy width, evident in Fig. \ref{fig2}(b). We trace the peak position across the island in Fig. \ref{figS4}(a) and compare it to the shift in the mid-position of the Coulomb blockade gap around zero bias (blue versus red lines, respectively). Evidently the two are perfectly anti-correlated up to a scaling prefactor of about 7 (notice the opposite and scaled corresponding voltage axes). The origin of the simultaneous shifts in both the peak and gap energies is the changes in the polarization charge accumulated on the island, $Q_0$, as the STM tip is rastered across it. Since the accumulated charge is positioned in between the voltage and drain contacts (biased tip and grounded nanowire, respectively) the electric field the accumulated charge induces is either added or deducted from the voltage drop across the two tunneling junctions, as noted by the opposite signs $Q_0$ enters with in Eq. \ref{eq2} and Eq. \ref{eq3}, respectively. As a result the Coulomb blockade gap and the Coulomb peaks shift in opposite directions in energy. The amounts of energy by which they shift is scaled by the corresponding capacitances, $C_{NA}$ and $C_{TA}$, as shown in Fig. \ref{fig3}(a). Remarkably, as the Coulomb peak shifts to lower energies towards the Coulomb blockade gap its width narrows, as shown in Fig. \ref{figS4}(b). We note that the energy width of the $dI/dV$ peak is larger than the instrumental broadening (temperature broadening is about $1 \ \mathrm{meV}$ and the AC excitation amplitude is 5 meV). This excess and energy dependent broadening is not accounted for in the semi-classical framework of the DBTJ model, and may originate from the finite dwell time of the electrons in the island,  which is determined by the larger tunneling rate $\Delta E \approx \hbar\Gamma_{NA}$ \cite{Millo2000}.
This trend signifies an increased interface resistance $R_{NA}$ at lower energies, as shown in Fig. \ref{figS4}(c), that might invalidate our assumption $R_{NA} \ll R_{TA}$. Indeed, at lower energies, below 200 meV, we find the most significant deviation of the measured $dI/dV$ from the simplified model used. 

We stress that the resistive properties of the InAs-Al junction we observe scale with the interface dimension. Since the exact dimensions of the larger droplets cannot be accurately inferred from the complex topography, we compare the scaling of $R_{NA}$ with the geometrical capacitive coupling across the interface $C_{NA}$. While the capacitance is naively proportional to the interface area $A$, the resistance is inversely proportional to it. We therefore expect to find a relation of the form $R_{NA} \sim C_{NA}^{-\alpha}$ with $\alpha \approx 1$ for a simple platelet geometry. Statistics of 10 different islands that stretch over an order of magnitude in resistance and capacitance (about 5-50 nm span in diameters) indeed find a simple power-law relation with $\alpha \approx 0.8$ (Fig. \ref{figS4}(d)) with no apparent deviation down to  the smallest islands inspected \cite{Zhang2018QuantumInAs}.

\section{Tip Induced Gating of the Charging Resonance }
\label{App:Off}

\begin{figure}
\includegraphics[width=0.5\linewidth]{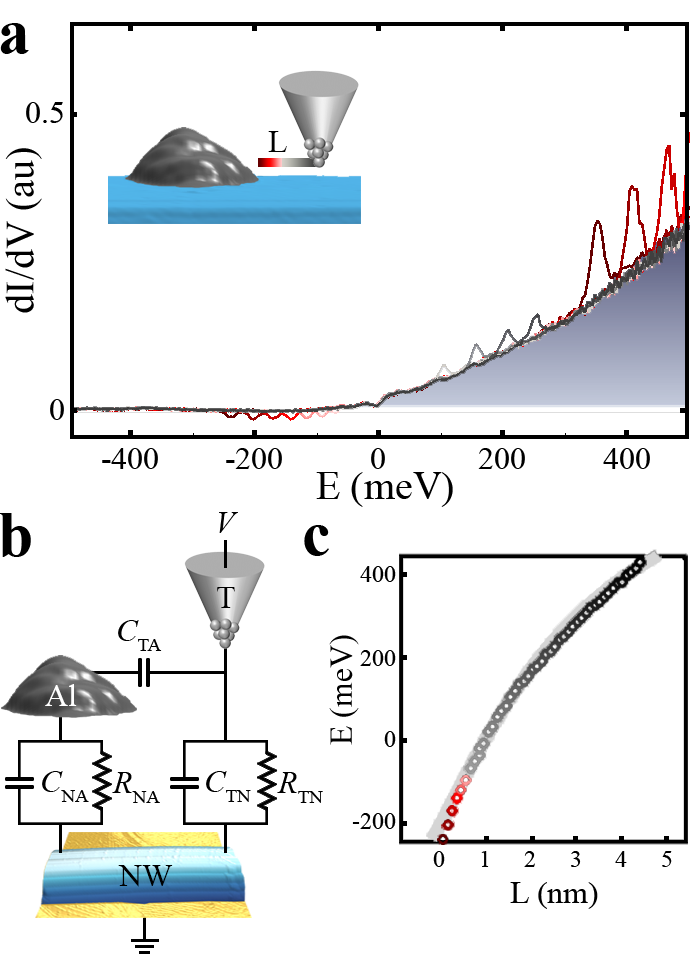}
\centering
\caption{\label{figS1} (a) $dI/dV$ curves at varying distance from the Al island, highlighting the resonance peaks. The red to gray color scale corresponds to the distance from the right boundary of the island as illustrated in the topography inset. The background signal
of the InAs spectrum is marked with solid blue. 
(b) A diagram of the corresponding circuit when the tip descends from the Al island
and becomes an effective gate electrode. (c) A trace following the
energy-distance dispersion of the resonance (circles) with a power-law fit (light gray) }
\end{figure}

In this section we discuss the possible origin of the two intriguing spectral features that appear on the nanowire in the vicinity of the Al island - the strong dispersion of the resonances and the negative differential conductance within the gap. Similar features where observed around islands on InAs(110) surfaces and could lead to local band bending \cite{Wildoer1996}, or be interpreted as coherent interference of electrons tunneling directly between the tip and the InAs and indirectly through the island \cite{Muzychenko2009}.  Fig. \ref{figS1}(a) shows a series of $dI/dV$ curves at varying distance from the island's right boundary (red to gray colors, as illustrated by the topography inset), the resonance peaks stand out with respect to the background of the InAs spectrum (solid blue). Note that the red colored resonances at higher energy emanate from the higher resonance of the island, whereas the NDC at negative energies emanate from the lower resonance (compare to Fig \ref{fig2}(b)). This resonance and its strong dispersion can be explained in the following manner; Once the tip descends off the Al island to the exposed InAs segment adjacent to it, the tip-Al tunneling path increases and $R_{TA}$ is suppressed exponentially over a distance $L \approx 1$ nm. Concurrently, a direct tunneling path between the tip and the nanowire with resistance $R_{TN}$ opens. However, even in the absence of a direct tunneling channel to the island a sharp Coulomb resonance as the one seen in Fig. \ref{fig2}(b) arises. Its occurrence off the island is guaranteed by the capacitive coupling between the tip and the Al island $C_{TA}$, that decays much slower than $R_{TA}$ - as a power-law rather than an exponent in $L$. The equivalent circuit for tunneling off the island is shown in Fig. \ref{figS1}(b). In this configuration the bias voltage applied to the capacitive gate $C_{TA}$ varies $Q_0$ on the island, and therefore allows electrons to jump on/off the island whenever $Q_0$ is tuned to $e(n_0 \pm 1/2)$. We can expand the distance dependence of $C_{TA}$ to second order in powers of $r/L$, with $r$ being an approximate diameter of the island (set to 2.5 nm) and $L$ is the distance to the tip.

\begin{equation} \label{CTA}
C_{TA}(L) = C_{TA}^0 (1+r/L+(r/L)^2)
\end{equation}

Plugging Eq. \eqref{CTA} to the expression for the charging energy $E_C(L) = e^2/2C(L)$, the observed dispersion can be reproduced with no further fitting parameters, As demonstrated by the fitted curve (light gray line) to a trace of the energy vs. distance dispersion of the resonance peak (colored circles) in Fig. \ref{figS1}(c).

Intriguingly, off the island and within the semiconducting gap the Coulomb peak shows negative differential conductance. In this regime there is no direct or indirect resistive coupling between the tip and the island because $R_{TN}, R_{NA} \rightarrow \infty$, as the nanowire itself turns insulating. Accordingly, all the couplings are capacitive, and the signal we measure is a pure displacement current. The negative reading results from the depletion of electrons beneath the tip that screen the positively charged adjacent island that can no longer be populated by additional electrons in the absence of any resistive coupling to the island.

\section{Derivation of a Generalized DBTJ Semi-classical Model} 

\label{App:Orthodox}

Following Refs. \cite{Hanna1991,Amman1989TheTransistor,Amman1991}, we model the system as a DBTJ, where each junction has capacitance $C_{NA}$ ($C_{TA}$) and resistance
$R_{NA}$ ($R_{TA}$) in parallel for the nanowire-aluminum (tip-aluminum) junction (refer to Fig. \ref{fig3}(b) for a diagram of the circuit). It should be noted that $R_{NA} \propto  (\left|T\right|^{2}\nu_{A}\nu_{N})^{-1}$, where $T$ is the tunneling matrix element and $\nu_{A}$ ($\nu_{N}$) is the density-of-states of the Al island (InAs nanowire). This will require a certain modification of the model, since $R_{NA}$ cannot be assumed to be energy independent, as often assumed in the framework of the DBTJ model. In our case $\nu_{N}$ and $T$ have a substantial energy dependence that should be accounted for.

The tunneling rates of any of the junctions connecting electrodes $a$ and $b$ can be calculated by using Fermi's Golden Rule:

\begin{align}
\Gamma_{a\rightarrow b} & \propto\int_{-\infty}^{\infty}\left|T\left(E-\varepsilon_{F}^{a}\right)\right|^{2}\nu_{a}\left(E-\varepsilon_{F}^{a}\right)\nu_{b}\left(E-\varepsilon_{F}^{b}\right)f\left(E-\varepsilon_{F}^{a}\right)\left(1-f\left(E-\varepsilon_{F}^{b}\right)\right)dE
\end{align}
where $f$ stands for the Fermi-Dirac distribution function and $\varepsilon_{F}$ is the Fermi energy. 

To calculate $\Gamma$, the semi-classical approximation $\varepsilon_{F}^{a}-\varepsilon_{F}^{b}=-\Delta E_{a\rightarrow b}^{class}$ is used. Namely, the difference between the Fermi energies of the electrodes comprising the junction corresponds to the difference between the energies of the classical system before and after an electron
has tunneled on/off the island:

\begin{align}
\Delta E_{TA}^{\pm} & =\frac{e}{C_{TA}+C_{NA}}\left(\frac{e}{2}\pm\left(N-\frac{Q_{0}}{e}\right)e\pm C_{NA}V\right)=E_{c}\left(1+2\left(N-\frac{Q_{0}}{e}\right)\right)\pm\frac{V}{1+\frac{C_{TA}}{C_{NA}}}\\
\Delta E_{NA}^{\pm} & =\frac{e}{C_{TA}+C_{NA}}\left(\frac{e}{2}\pm\left(N-\frac{Q_{0}}{e}\right)e\mp C_{TA}V\right)=E_{c}\left(1-2\left(N-\frac{Q_{0}}{e}\right)\right)\mp\frac{V}{1+\frac{C_{NA}}{C_{TA}}}
\end{align}

Where $\pm$ denotes tunneling on/off the central electrode (the Al island), $E_{c}\equiv e^2/ 2\left(C_{TA}+C_{NA}\right)$
is the charging energy, $Q_{0}$ is the polarization charge, $N$ is
the number of electrons on the central electrode (island charge), assumed zero at zero bias, and $V$ is the voltage bias.

The temperature scale in the experiment is $\sim 1$ meV, much smaller than all the other relevant scales, and is therefore treated as zero temperature.

Combining the above equations we obtain:

\begin{align} \label{TunnelingRate}
\Gamma_{a\rightarrow b} & \propto\int_{-\infty}^{\infty}\left|T\left(E-\varepsilon_{F}^{a}\right)\right|^{2}\nu_{a}\left(E-\varepsilon_{F}^{a}\right)\nu_{b}\left(E-\varepsilon_{F}^{b}\right)f\left(E-\varepsilon_{F}^{a}\right)\left(1-f\left(E-\varepsilon_{F}^{b}\right)\right)dE\\ \nonumber
 & =\int_{-\infty}^{\infty}\left|T\left(E\right)\right|^{2}\nu_{a}\left(E\right)\nu_{b}\left(E+\varepsilon_{F}^{a}-\varepsilon_{F}^{b}\right)f\left(E\right)\left(1-f\left(E+\varepsilon_{F}^{a}-\varepsilon_{F}^{b}\right)\right)dE\\ \nonumber
 & \overset{\varepsilon_{F}^{a}-\varepsilon_{F}^{b}=-\Delta E_{a\rightarrow b}^{class}}{=}\int_{-\infty}^{\infty}\left|T\left(E\right)\right|^{2}\nu_{a}\left(E\right)\nu_{b}\left(E-\Delta E_{a\rightarrow b}^{class}\right)f\left(E\right)\left(1-f\left(E-\Delta E_{a\rightarrow b}^{class}\right)\right)dE\\ \nonumber
 & \overset{T\rightarrow0}{=}\int_{-\infty}^{\infty}\left|T\left(E\right)\right|^{2}\nu_{a}\left(E\right)\nu_{b}\left(E-\Delta E_{a\rightarrow b}^{class}\right)\left(1-\Theta\left(E\right)\right)\Theta\left(E-\Delta E_{a\rightarrow b}^{class}\right)dE\\ \nonumber
 & =\Theta\left(-\Delta E_{a\rightarrow b}^{class}\right)\int_{\Delta E_{a\rightarrow b}^{class}}^{0}\left|T\left(E\right)\right|^{2}\nu_{a}\left(E\right)\nu_{b}\left(E-\Delta E_{a\rightarrow b}^{class}\right)dE
\end{align}
here $\Theta(E)$ is the Heaviside step function which substitutes $1-f$ in the zero temperature limit.

We define $\sigma\left(N,V\right)$ as the classical probability distribution of having $N$ electrons at bias $V$. Noting that the proportionality constant is the same for all tunneling rates, we need to solve the detailed balance equation
subject to the probability normalization:

\begin{align}
\sigma\left(N,V\right)\underset{\begin{array}{c}
\text{total tunneling rate into central}\\
\text{ electrode with \ensuremath{N} electrons}
\end{array}}{\underbrace{\left(\Gamma_{+}^{TA}\left(N,V\right)+\Gamma_{+}^{NA}\left(N,V\right)\right)}}=\ \underset{\begin{array}{c}
\text{total tunneling rate out of central}\\
\text{ electrode with \ensuremath{N+1} electrons}
\end{array}}{\underbrace{\left(\Gamma_{-}^{TA}\left(N+1,V\right)+\Gamma_{-}^{NA}\left(N+1,V\right)\right)}}\sigma\left(N+1,V\right)
\end{align}

\begin{align}
\sum_{N=-\infty}^{\infty}\sigma\left(N\right)=1
\end{align}

This equation has an analytic solution \cite{Amman1991} that is given by:

\begin{align}
\sigma\left(N,V\right) & =\frac{\prod_{i=-\infty}^{N-1}\left(\Gamma_{+}^{TA}\left(i,V\right)+\Gamma_{+}^{NA}\left(i,V\right)\right)\prod_{i=N+1}^{\infty}\left(\Gamma_{-}^{TA}\left(i,V\right)+\Gamma_{-}^{NA}\left(i,V\right)\right)}{\sum_{j=-\infty}^{\infty}\left[\prod_{i=-\infty}^{j-1}\left(\Gamma_{+}^{TA}\left(i,V\right)+\Gamma_{+}^{NA}\left(i,V\right)\right)\prod_{i=j+1}^{\infty}\left(\Gamma_{-}^{TA}\left(i,V\right)+\Gamma_{-}^{NA}\left(i,V\right)\right)\right]}
\end{align}

However, practical evaluation of this expression requires truncating the sum for larger values of $\left|N\right|$. This approximation is based on the fact that $\sigma(N,V)$ is sharply peaked at low temperature and reasonable voltages around the most probable occupation $N_0$, which is of order 1.  The current (up to a proportionality constant) is then just the difference
between the tunneling rates in and out of any of the junctions:
\begin{align}
\label{DBTJ_current}
I\left(V\right) & \propto e\sum_{N=-\infty}^{\infty}\sigma\left(N\right)\left(\Gamma_{+}^{NA}\left(N\right)-\Gamma_{-}^{NA}\left(N\right)\right)\\ \nonumber
 & \propto e\sum_{N=-\infty}^{\infty}\sigma\left(N\right)\left(\Gamma_{-}^{TA}\left(N\right)-\Gamma_{+}^{TA}\left(N\right)\right)
\end{align}
And $dI/dV$ is obtained by numerical differentiation with respect to $V$. To account for the transmission probability $T(E)$ and for an energy-dependent $\nu$ in the superconductor case, we simply assign them explicit expressions, as described in the main text, and evaluate Eq. \eqref{TunnelingRate} numerically.

\section{Spectroscopy of gated nanowires}
\label{App:gating}

\begin{figure}
\includegraphics[width=1\linewidth]{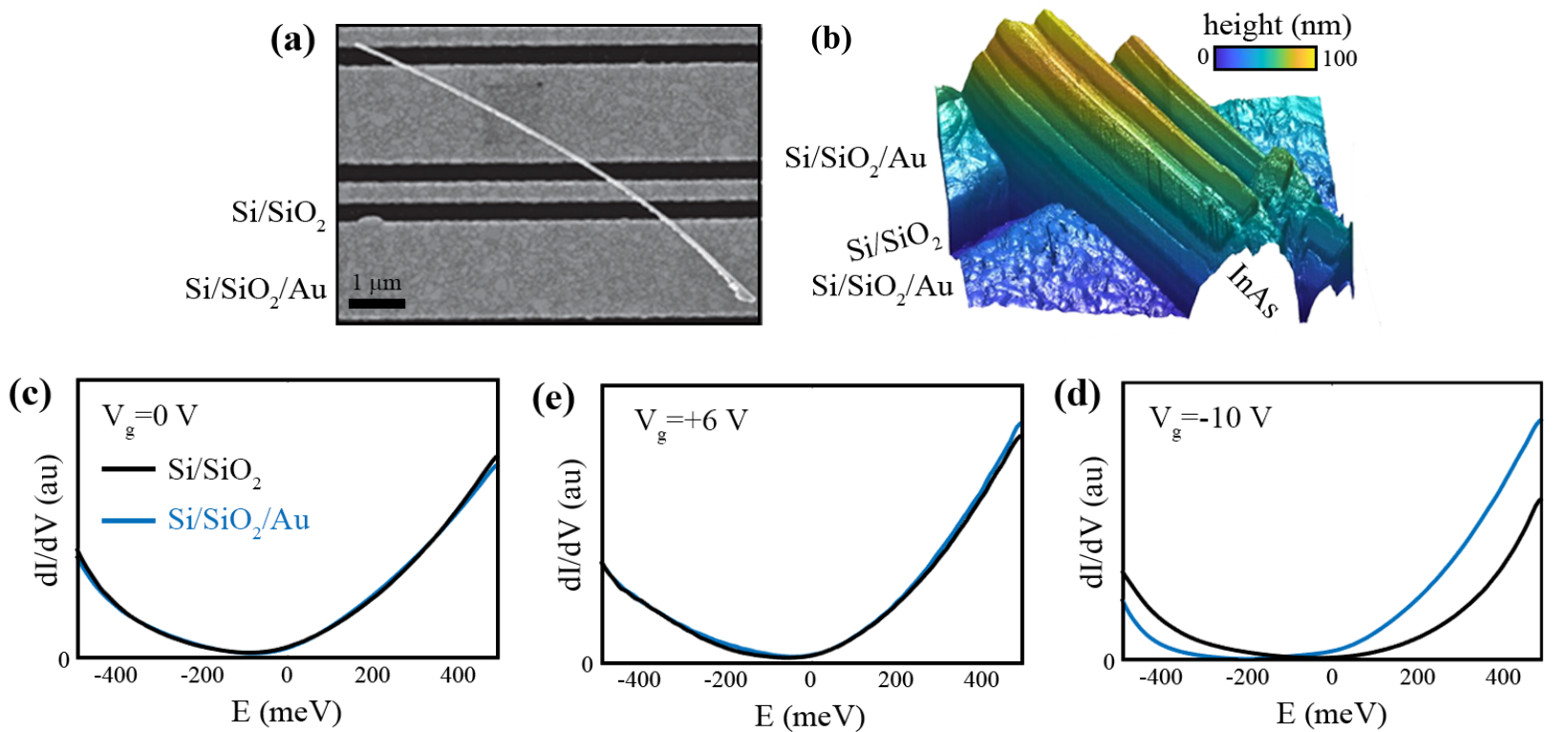} 
\centering
\caption{\label{figS3} (a) SEM image of a gatable nanowire device for STM measurement. 100 nm Au film is deposited over doped Si/SiO$_2$ substrate. Focused ion beam is then used to remove Au along 200 nm wide strips exposing the SiO$_2$ layer beneath. Nanowires are then deposited \textit{in-situ}. (b) Topographic image obtained in a STM  operated at room temperature of a nanowire across a carved trench in the Si/SiO$_2$/Au substrate. (c) At $V_g=0$ we find no change in the $dI/dV$ spectrum of the the nanowire segment that rests on Au to that suspended over SiO$_2$ (blue versus black, respectively). (d) In response to application of $V_g=-10 V$ the $dI/dV$ spectrum of the suspended segment of the nanowire is shifted by about 100 meV with respect to that measured over a segment resting on Au. To remove the set-point effect, in which the STM feedback maintains a constant integrated LDOS between zero and parking bias by adjusting the tip to sample height, the spectra are normalized by the integrated LDOS in the 300 meV energy window above the onset of the conduction band. (e) No significant response in the $dI/dV$ to application of positive bias of $V_g=+6V$ due to Fermi level pinning.  }  
\end{figure}

We provide preliminary results of spectroscopic STM measurements of capacitively gated nanowires. While gating necessitates electrical isolation between the gate electrode and the mesoscopic object to be gated, STM requires to drain the tunneling electrons which is commonly obtained by using metallic substrates. Gatable nanowire devices suitable for STM measurements thus require an additional technological solution. We have fabricated suitable substrates by exposing narrow slits in a Au deposited Si/SiO$_2$ chips using a focused ion beam. We then harvest and deposit InAs nanowires \textit{in-situ} similar to the way we do with Au single crystals substrates. Some of the deposited nanowires happen to rest across exposed slits, as shown in the SEM image in Fig. \ref{figS3}(a). These nanowire segment should then be gatable. A topographic STM image of a nanowire residing over an exposed slit in the Au substrate measured with a home-built STM is shown in Fig. \ref{figS3}(b). The $dI/dV$ spectra in Fig. \ref{figS3}(c), measured on the InAs nanowire in a region that rests on Au in topography and in a region in which the nanowire is suspended over a slit (blue and black lines, respectively), are identical. This exemplifies again the robustness of the $dI/dV$ spectrum of InAs nanowires to the introduction of metallic contacts due to Fermi level pinning. However, under application of a -10 V gate voltage we find a clear energy shift in the $dI/dV$ spectrum between the suspended and non-suspended segments, shown in Fig. \ref{figS3}(d). The former is shifted by about 100 meV from the latter. This shows that over the exposed slits we are able to almost deplete the nanowire from electrons while elsewhere the metallic Au film screens the effect of the back-gate. In contrast, we do not find similar response to application of positive gate voltage, as demonstrated in Fig. \ref{figS3}(e). The nanowires can be depleted by a negative gate voltages, additional charges can hardly be accumulated by an applied positive gate voltage. We attribute this rectification in the response of the nanowire to gating to Fermi level pinning. Once an electron density of about 10$^{12}$ cm$^{-2}$ is achieved additional charges are hardly added. However, by application of sufficient negative gate voltages the Fermi level can be depinned. Accordingly, Fermi level pinning, induced by the dense surface states in the InAs nanowires, allows only to effectively deplete the electrons from the nanowires.


%





\bibliography{Mendeley_AlInAs}

\end{document}